\documentclass[runningheads]{llncs}
\usepackage{lmodern}
\usepackage{amssymb,amsmath}
\usepackage{ifxetex,ifluatex}
\usepackage{fixltx2e} 
\ifnum 0\ifxetex 1\fi\ifluatex 1\fi=0 
  \usepackage[T1]{fontenc}
  \usepackage[utf8]{inputenc}
\else 
  \ifxetex
    \usepackage{mathspec}
  \else
    \usepackage{fontspec}
  \fi
  \defaultfontfeatures{Ligatures=TeX,Scale=MatchLowercase}
  
\fi
\IfFileExists{upquote.sty}{\usepackage{upquote}}{}
\IfFileExists{microtype.sty}{%
\usepackage{microtype}
\UseMicrotypeSet[protrusion]{basicmath} 
}{}
\usepackage{hyperref}
\PassOptionsToPackage{usenames,dvipsnames}{color} 
\hypersetup{unicode=true,
            pdftitle={Monte Carlo Tableau Proof Search},
            pdfauthor={Michael Färber; Cezary Kaliszyk; Josef Urban},
            pdfborder={0 0 0},
            breaklinks=true}
\usepackage{longtable,booktabs}
\providecommand{\tightlist}{%
  \setlength{\itemsep}{0pt}\setlength{\parskip}{0pt}}
\title{Monte Carlo Tableau Proof Search}
\author{Michael Färber\inst{1} \and Cezary Kaliszyk\inst{1} \and Josef Urban\inst{2}}

\institute{Universität Innsbruck, Austria
\linebreak \email{\{michael.faerber,cezary.kaliszyk\}@uibk.ac.at} \and Czech Technical University in Prague, Czech Republic
\linebreak \email{josef.urban@gmail.com}}

\date{}
\usepackage{centerfloat}
\usepackage{tikz-qtree}
\usepackage[svgpath=svg/]{svg}

\usepackage{subfig}

\usepackage{pgfplots}
\pgfplotsset{compat=1.9}
\usepgfplotslibrary{units}
\usepackage{booktabs}

\usepackage{algorithm}
\usepackage[noend]{algpseudocode}
\DeclareMathOperator*{\argmax}{arg\,max}

\ifx\paragraph\undefined\else
\let\oldparagraph\paragraph
\renewcommand{\paragraph}[1]{\oldparagraph{#1}\mbox{}}
\fi
\ifx\subparagraph\undefined\else
\let\oldsubparagraph\subparagraph
\renewcommand{\subparagraph}[1]{\oldsubparagraph{#1}\mbox{}}
\fi

\begin{document}
\maketitle
\begin{abstract}
We study Monte Carlo Tree Search to guide proof search in tableau
calculi. This includes proposing a number of proof-state evaluation
heuristics, some of which are learnt from previous proofs. We present an
implementation based on the leanCoP prover. The system is trained and
evaluated on a large suite of related problems coming from the Mizar
proof assistant, showing that it is capable to find new and different
proofs.
\end{abstract}

\hypertarget{introduction}{%
\section{Introduction}\label{introduction}}

Recent advances in Automated Reasoning include both theoretical
improvements in the calculi, including combining superposition with SAT
solving in recent versions of Vampire \cite{biere2014-satvampire} and
research on the InstGen calculus in iProver \cite{korovin2013-instgen},
but also more practical improvements, such as more efficient and precise
term indexing techniques \cite{schulz2013-e}, efficient non-clausal
tableau proof search \cite{otten2016-nanocop}, or the use of machine
learning for problem size reduction \cite{kaliszyk2015-et}.
Furthermore, many automated reasoning techniques have been extended to
interesting theories beyond first-order logic, including the
developments in SMT solving in CVC4 \cite{barrett2011-cvc4} or to
higher-order logic \cite{wisniewski2015-leopard,brown2012-satallax}.
Many of these developments have been of great value for interactive
theorem provers, whose most powerful general purpose automation
techniques today rely on automated reasoning tools
\cite{blanchette2016-qed}.

However, current automated theorem provers are still quite weak in
finding more complicated proofs, especially over large formal
developments \cite{urban2010-evaluation}. The search typically blows up
after several seconds, making the chance of finding proofs in longer
times exponentially decreasing \cite{alama2012-comparison}. This
behaviour is reminiscent of poorly guided search in games such as chess
and Go. The number of all possible variants there typically also grows
exponentially, and intelligent guiding methods are needed to focus on
exploring the most promising moves and positions.

The guiding method that has recently very significantly improved
automatic game play is Monte Carlo Tree Search (MCTS), i.e., expanding
the search based on its (variously guided) random sampling
\cite{browne2012-survey}. Recent developments in MCTS include
combination of exploration and exploitation \cite{kocsis2006-uct},
combination of online and offline knowledge with the All-Moves-As-First
(AMAF) heuristic \cite{gelly2007-uct}, and adaptive tuning of rollout
policies during search \cite{rosin2011-rollout}. As shown for example
in the AlphaGo system \cite{silver2016-alphago}, machine learning can
be used to train good position evaluation heuristics even in very
complicated domains that were previously thought to be solely in the
realm of ``human intuition''. From the point of game theory, automated
theorem proving is a combinatorial single-player game. For some games in
this category, including SameGame \cite{schadd2012-samegame} and the
NP-hard Morpion Solitaire \cite{rosin2011-rollout}, MCTS has produced
state-of-the-art players. \cite{hoder2016-selection} shows that proof
search can be positively guided by one-step lookahead, and MCTS allows
approximation of multi-step lookaheads by use of random sampling. While
``finishing the randomly sampled game'' -- as used in the most
straightforward MCTS for games -- is not always possible in ATP (it
would mean finishing the proof), there is a chance of learning good
\emph{proof state evaluation heuristics} that will guide MCTS for ATPs
in a similar way as e.g.~in AlphaGo.

In this work, we study MCTS methods that can guide the search in
automated theorem provers, and evaluate their impact on interactive
theorem proving problems in first-order logic. We focus on the tableau
calculus and on the leanCoP prover \cite{otten2008-leancop}, which has
a compact implementation that is easy to experiment with. We can also
build on previous machine learning extensions of leanCoP
\cite{kaliszyk2015-femalecop,urban2011-malecop}. To our knowledge,
this is the first time MCTS has been applied to theorem proving.

\hypertarget{contributions}{%
\subsection*{Contributions}\label{contributions}}
\addcontentsline{toc}{subsection}{Contributions}

We introduce a set of MCTS heuristics tailored to proof search including
two state transition probability heuristics, three state evaluation
heuristics, and two tree expansion policies related to restricted
backtracking (\autoref{tableau-heuristics}). Furthermore, we present an
implementation interleaving a traditional proof search with MCTS
(\autoref{implementation}) and measure its performance on a set of Mizar
Mathematical Library problems (\autoref{evaluation}).

\hypertarget{mcts}{%
\section{Monte Carlo Tree Search}\label{mcts}}

Monte Carlo Tree Search (MCTS) is a technique to guide search in a large
decision space by taking random samples and evaluating their outcome.
First, we will establish a format for problems tractable with MCTS.
Then, we give a notation for Monte Carlo trees. Finally, we show how to
create and evolve a Monte Carlo tree for given problems with
problem-specific heuristics.

\hypertarget{problem-setting}{%
\subsection{Problem Setting and Example}\label{problem-setting}}

A tree search problem can be minimally characterised with:

\begin{itemize}
\tightlist
\item
  a set of states \(\mathcal{S}\),
\item
  an initial state \(s_0 \in \mathcal{S}\), and
\item
  a state transition function
  \(\delta : \mathcal{S} \to 2^\mathcal{S}\).
\end{itemize}

As an example of a tree search problem, consider the travelling salesman
problem: A salesman has to visit a set of cities \(C\) and wants to
minimise the total distance travelled, where \(d(c_1, c_2)\) is the
distance between two cities.

A possible tree search characterisation of the travelling salesman
problem is:

\begin{itemize}
\tightlist
\item
  The set of states \(\mathcal{S}\) are the sequences of cities visited
  so far; for example {[}Prague{]}, {[}Prague, Vienna, Bratislava{]},
  {[}Paris{]}.
\item
  The initial state \(s_0\) is the empty sequence.
\item
  The state transition function \(\delta\) returns for a sequence of
  already visited cities the set of sequences where one previously
  unvisited city is added; i.e.
  \(\delta(s) = \bigcup _ {c \in C, c \notin s} [s,c]\). For example,
  \(\delta([\text{Prague, Vienna}])\) could contain {[}Prague, Vienna,
  Budapest{]} and {[}Prague, Vienna, Bratislava{]}.
\end{itemize}

The number of states of the travelling salesman problem is exponential
in the number of cities, therefore constructing a whole tree to obtain
an optimal solution is not feasible. To bias the tree search towards
more promising regions, we define two types of heuristics:

\begin{itemize}
\tightlist
\item
  the probability \(P : \mathcal{S} \to [0,1]\) of choosing a state and
\item
  the reward \(\rho : \mathcal{S} \to [0,1]\) for a state.
\end{itemize}

\(P(s' \mid s)\) is the probability for choosing state \(s'\) when being
in its predecessor state \(s\), where \(s' \in \delta(s)\). In the
travelling salesman example,
\[P([\text{Prague, Vienna}] \mid [\text{Prague}]) >
 P([\text{Prague, Paris}]  \mid [\text{Prague}])\] means that when the
salesman is in Prague, Vienna should be chosen as next city with a
higher probability than Paris.

\(\rho(s)\) is the overall quality of a (final) state \(s\). Due to the
MCTS flavour we use \cite{kocsis2006-uct}, it has to be normed between
{[}0,1{]}. In the travelling salesman example, a sensible \(\rho\)
should yield larger values for city sequences with smaller overall
distance.

\hypertarget{trees}{%
\subsection{Trees}\label{trees}}

A Monte Carlo tree stores the states that have been expanded during a
tree search, and keeps statistics about the states. We define the set
\(\mathcal{T}\) of Monte Carlo tree nodes.

\begin{definition}[Monte Carlo tree node]
A \emph{Monte Carlo tree node} is a 5-tuple $(n, r, s, S, T) \in \mathcal{T}$, where:

\begin{itemize}
\item $n \in \mathbb{N}$ is the number of times the node was visited,
\item $r \in \mathbb{R}$ is the sum of the rewards of successors,
\item $s \in \mathcal{S}$ is the state of the node,
\item $S \in 2^\mathcal{S}$ are the unvisited successor states of $s$, and
\item $T \in 2^\mathcal{T}$ are the child tree nodes.
\end{itemize}
\end{definition}

Monte Carlo Tree Search evolves a tree by repeatedly applying a step
function until a certain criterion is fulfilled, e.g.~a certain number
of steps is performed, time has elapsed etc. Ideally, every step should
refine the quality estimate of the states in the Monte Carlo tree. We
show the step function in the next section.

\hypertarget{mcsf}{%
\subsection{Monte Carlo Step Function}\label{mcsf}}

The Monte Carlo step function performs the following: First, it selects
a \(node\) in the Monte Carlo tree. From the state of \(node\), it
randomly samples a sequence of successor states (called
\emph{simulation}). It then creates a new \(node'\) with some state from
the simulation and makes it a child of the original \(node\). Finally, a
reward is calculated from the simulation and backpropagated to all
ancestors of \(node'\).

The idea is that rewards obtained from simulations starting from a
certain node let us estimate the usefulness of the node itself. A
description of the pseudocode in algorithm \ref {alg:mcts-step} follows.
For brevity, the pseudocode assumes that every state has at least one
successor state (i.e., for every \(s \in \mathcal{S}\),
\(\left\vert \delta(s) \right\vert > 0\)).

\begin{algorithm}
\caption{Monte Carlo step function.}\label{alg:mcts-step}
\begin{algorithmic}[1]
\Procedure{step}{$node$}
\If {$node.S = \emptyset$}
  \State $best \gets \argmax _ {t \in node.T} \Call{uct}{node.n, t}$
    \Comment{selection}
  \State $reward \gets \Call{step}{best}$
\Else
  \State $s' \gets \Call{biasedDraw}{P, node.s, node.S}$
  \State $sim \gets \Call{simulation}{D, s'}$
    \Comment{simulation}
  \State $(node', reward) \gets \Call{expansion}{sim}$
    \Comment{expansion}
  \State $node.S \gets node.S \setminus \left\{s'\right\}$
  \State $node.T \gets node.T \cup \left\{node'\right\}$
\EndIf
\State $node.n \gets node.n + 1$ \Comment{backpropagation}
\State $node.r \gets node.r + reward$
\State \Return{reward}
\EndProcedure
\end{algorithmic}
\end{algorithm}

In l. 3, the step function recursively selects the child node with the
highest UCT (Upper Confidence Bounds for Trees) value
\cite{kocsis2006-uct}. UCT establishes an order on nodes, combining
\emph{exploration} and \emph{exploitation}: \emph{exploration} prefers
less frequently visited nodes, whereas \emph{exploitation} prefers nodes
with higher average reward. The ratio between these two goals is
determined by the exploration constant \(C_p\), where higher values give
more emphasis to exploration. The average reward of a node
\((n_j, r_j, s, S, T)\) is \(\frac{r_j}{n_j}\). The UCT function takes
the number of times \(n\) that a parent node was visited, as well as a
child node:
\[\text{uct}(n, (n_j, r_j, s, S, T)) = \frac{r_j}{n_j} + C_p \sqrt\frac{\ln n}{n_j}.\]

As soon as a node with unvisited successor states is encountered (l. 5),
an unvisited successor state \(s'\) is drawn (l. 6), where the
probability of picking \(s'\) is proportional to \(P(s' \mid s)\). From
the chosen \(s'\), a simulation is performed up to a constant simulation
depth \(D\). A simulation starting from a state \(s_i\) draws a state
\(s _ {i+1}\) from \(\delta(s_i)\), with probability
\(P(s _ {i+1} \mid s_i)\). This is repeated for a certain number of
times, yielding a simulation \(\left[s_1, \dots, s_D\right]\), where
\(D\) is the simulation depth and for every \(i\),
\(s _ {i+1} \in \delta(s_i)\).

From the simulation, the expansion operation yields a new \(node'\) and
a \(reward\). The default expansion policy creates \(node'\) from the
first state \(s_1\) of the simulation and calculates the reward from the
last state, i.e. \(reward = \rho(s_D)\). Therefore, the new expansion
node is \(node' = (1, \rho(s_D), s_1, \delta(s_1), \emptyset)\).

The expansion node \(node'\) is added to the child nodes (l. 10) and the
\(reward\) is propagated back until the root (l. 11-13).

\hypertarget{tableau}{%
\section{Tableau}\label{tableau}}

In this section, we shortly recall some basics of tableau
\cite{haehnle2001-tableaux} and introduce notions specific to
representing tableau as MCTS.

Tableau calculi are methods to prove the inconsistency of formulae. A
tableau is a tree with formulae as nodes. The root is the formula whose
inconsistency one attempts to show. All other nodes are produced by
application of tableau rules to nodes above them; such rules include
\(\alpha\)-rules to treat conjunctions and \(\beta\)-rules to treat
disjunctions. For example, when a branch contains a disjunction (called
\(\beta\)-formula), then an application of a \(\beta\)-rule
(parametrised by the disjunction) adds the disjuncts as children to some
leaf of the branch.

The choice of \(\beta\)-formulae in tableau proof search is one of the
main sources of nondeterminism and has a considerable impact on the
length of the proof search. It corresponds to the choice of given
clauses in saturation-based provers and extension clauses in the
connection calculus. Therefore, in this work, we focus on influencing
the proof search mostly by influencing the choice of \(\beta\)-formulae.
We will abstract from the actual tableau steps, only assuming that the
considered tableau calculus is sound and complete.

A branch of a tableau is closed iff it contains some formula and its
negation. A tableau is closed iff all of its branches are closed. A
formula is proven inconsistent when there is a closed tableau with the
formula at the root. For the heuristics in \autoref{tableau-heuristics},
we define \emph{\(\beta\)-children}, which correspond to \emph{open
subgoals} in interactive theorem provers and literals of \emph{open
branches} in the connection calculus.

\begin{definition}[$\beta$-children]
Given a tableau $t$, we call direct children of branches
\emph{$\beta$-children} of $t$ and denote them as $\beta(t)$.

\emph{Open $\beta$-children} (denoted $\beta _ o(t)$) are
all $\beta$-children on open branches not having any branch as descendant.
\emph{Closed $\beta$-children} (denoted $\beta _ c(t)$) are
all $\beta$-children that are not open,
i.e. $\beta _ c(t) = \beta(t) \setminus \beta_o(t)$.
\end{definition}

\begin{figure}
\subfloat[State 1.]{
  \begin{tikzpicture}[
  connect/.style={semithick,<->},
  every label/.style={draw, circle, inner sep = 1 pt}]
  \Tree
  [.{$f$}
   [.\framebox{$p$} ]
   [.\framebox{$q$} ]
   [.\framebox{$r$} ]
  ]
  \end{tikzpicture}
}
\hfill
\subfloat[State 2.]{
  \begin{tikzpicture}[
  connect/.style={semithick,<->},
  every label/.style={draw, circle, inner sep = 1 pt}]
  \Tree
  [.{$f$}
   [.\node (p) {$p$};
    [.\node (np) {$\lnot p$}; ]
    [.\framebox{$s$} ]
   ]
   [.\framebox{$q$} ]
   [.\framebox{$r$} ]
  ]
  \draw[connect] (p.west) [bend right] to (np.west);
  \end{tikzpicture}
}
\hfill
\subfloat[State 3.]{
  \begin{tikzpicture}[
  connect/.style={semithick,<->},
  every label/.style={draw, circle, inner sep = 1 pt}]
  \Tree
  [.{$f$}
   [.\node (p) {$p$};
    [.\node (np) {$\lnot p$}; ]
    [.\node (s) {$s$};
     [.\node (ns) {$\lnot s$}; ]
    ]
   ]
   [.\framebox{$q$} ]
   [.\framebox{$r$} ]
  ]
  \draw[connect] (p.west) [bend right] to (np.west);
  \draw[connect] (s.east) [bend left] to (ns.east);
  \end{tikzpicture}
}
\hfill
\subfloat[State 4.]{
  \begin{tikzpicture}[
  connect/.style={semithick,<->},
  every label/.style={draw, circle, inner sep = 1 pt}]
  \Tree
  [.{$f$}
   [.\node (p) {$p$};
    [.\node (np) {$\lnot p$}; ]
    [.\node (s) {$s$};
     [.\node (ns) {$\lnot s$}; ]
    ]
   ]
   [.\node (q) {$q$};
    [.\node (nq) {$\lnot q$}; ]
    [.\node (t) {\framebox{$t$}}; ]
   ]
   [.{\framebox{$r$}} ]
  ]
  \draw[connect] (p.west) [bend right] to (np.west);
  \draw[connect] (s.east) [bend left] to (ns.east);
  \draw[connect] (q.west) [bend right] to (nq.north west);
  \end{tikzpicture}
}
\caption{Proof search for formula $f =
  (p \lor q \lor r) \land (\lnot p \lor s) \land (\lnot p \lor t \lor u) \land
  \lnot s \land (\lnot q \lor t) \land (\lnot q \lor s)$.
  Open $\beta$-children are surrounded by boxes.}
\label{fig:proof-search}
\end{figure}

\begin{example}
In state 4 in \autoref{fig:proof-search}, closed $\beta$-children are
$p$, $\lnot p$, $s$, $q$, and $\lnot q$.
Open $\beta$-children are $t$ and $r$.
\end{example}

The \emph{successor tableaux} of a tableau are all tableaux that can be
obtained from the original tableau by the application of some rule. We
now give a description of tableau construction for a given formula \(f\)
in the language of \autoref{problem-setting}:

\begin{itemize}
\tightlist
\item
  The set of states \(\mathcal{S}\) is the set of tableaux.
\item
  The initial state \(s_0\) is a tableau containing only the formula
  \(f\) as root.
\item
  The transition function \(\delta(s)\) obtains all successor tableaux
  of \(s\) produced by applications of tableaux rules.
\end{itemize}

This characterisation in conjunction with the default expansion policy
from \autoref{mcsf} has the downside that its Monte Carlo trees are
approximately as deep as the number of proof \emph{steps}, whereas the
corresponding tableaux are as deep as the maximal proof \emph{depth}.
For example, the TPTP \cite{sutcliffe2013-j6} problem \texttt{PUZ035-1}
permits a proof consisting of about 40 proof steps in a tableau of depth
6. The Monte Carlo tableau characterisation, however, requires building
a Monte Carlo search tree with a depth close to 40, which is challenging
even when using a good state reward \(\rho\). The required tree depth
can be often decreased with the tableau-specific expansion policies
described in \autoref{beta-min-expansion}, but finding a
characterisation that reliably reduces the depth of the search tree
remains future work.

\hypertarget{tableau-heuristics}{%
\section{Tableau Heuristics}\label{tableau-heuristics}}

In \autoref{problem-setting}, we defined two kinds of heuristics to
guide Monte Carlo Tree Search, namely transition probability and state
reward. In this section, we propose such heuristics, as well as a set of
incomplete expansion policies.

\hypertarget{transition-probability}{%
\subsection{Transition Probability}\label{transition-probability}}

The transition probability \(P(s' \mid s)\) is the probability of
choosing state \(s'\) as successor state when in state \(s\), where
\(s' \in \delta(s)\). \(P\) is used to bias the selection of a successor
state in random simulations, as well as to determine the order of
visiting previously unvisited successor states; see algorithm
\ref{alg:mcts-step}.

When in some state \(s\), different kinds of tableau rules might be
applicable; for example \(\alpha\)-rules and \(\beta\)-rules (similarly
to extension and reduction rules in the connection calculus). In this
work, we focus on influencing the probability of \(\beta\)-rules
depending on their used \(\beta\)-formulae, which corresponds to earlier
work about choosing good extension clauses in the connection calculus
\cite{urban2011-malecop}. Therefore, we only vary the probabilities of
\(\beta\)-rules and attribute to all non-\(\beta\)-rules the same
probabilities.

As transition probabilities are among of the most frequently calculated
values in Monte Carlo Tree Search, the speed of this heuristic is
important. The baseline heuristic is to give the same probability to all
transitions, i.e. \(P_1(s' \mid s) \propto 1.\)

\hypertarget{beta-size}{%
\subsubsection{\texorpdfstring{\(\beta\)-size}{\textbackslash{}beta-size}}\label{beta-size}}

The \(\beta\)-size heuristic attributes a probability to a
\(\beta\)-rule that is inversely proportional to the number of newly
opened \(\beta\)-children:
\[P _ \beta(s' \mid s) \propto (\left|\beta _ o(s')\right| - \left|\beta _ o(s)\right|)^{-1}.\]

\begin{example}
In state 1 of \autoref{fig:proof-search},
it is possible to apply the $\beta$-rule to the leftmost branch with either
$\lnot p \lor s$ or $\lnot p \lor t \lor u$.
The first formula consists of two disjuncts and the second
of three disjuncts, so
the $\beta$-size heuristic attributes a probability proportional to
$\frac{1}{2}$ to the first and $\frac{1}{3}$ to the second formula.
The probabilities are normalized to sum to $1$, obtaining the actual
values $\frac{3}{5}$ and $\frac{2}{5}$ respectively.
\end{example}

\hypertarget{naive-bayes}{%
\subsubsection{Naive Bayesian Probability}\label{naive-bayes}}

Given the information about the formulae that were used in previous
successful proofs at particular proof states, it is possible to
calculate the likelihood that a given formula contributes to the current
proof attempt in the current proof state. Naive Bayesian probability is
used in \cite{kaliszyk2015-femalecop} to order formulae by
\[P(l_i \mid \vec f) = \frac{P(l_i) P(\vec f \mid l_i)}{P(\vec f)} \propto
  P(l_i) \prod_j P(f_j \mid l_i),\] where \(l_i\) is a \(\beta\)-formula
from a set \(\vec l\) of applicable \(\beta\)-formulae, and \(\vec f\)
is a set of features that characterises the current tableau, such as its
formulae symbols.

\(P(l_i)\) and \(P(f_j \mid l_i)\) as in \cite{kaliszyk2015-femalecop}
frequently yield values such that the probability of applying
\(\beta\)-rules is magnitudes smaller than for non-\(\beta\)-rules,
slowing down proof search. For that reason, we introduce normed
probability estimates.

First, let us denote the knowledge about the usage of \(\beta\)-formulae
in previous proofs by \(F(l_i)\), which is the multiset of sets of
features having occurred in conjunction with \(l_i\) when \(l_i\) was
used in a proof. \(\left| F(l_i) \right|\) is the total number of times
that \(l_i\) was used in previous proofs.

\begin{example}
$F(l_1) = \left\{\left\{f_1, f_2\right\}, \left\{f_2, f_3\right\}\right\}$
means that the formula $l_1$ was used twice in previous proofs;
once in a situation characterised by the features $f_1$ and $f_2$, and
once when features $f_2$ and $f_3$ were present.
\end{example}

This allows us to write the normed formula probability as
\[P(l_i) = \frac{\left| F(l_i) \right|}{\max _ {l_j \in \vec l} \left| F(l_j) \right|}.\]
Using \(\max\) instead of \(\sum\) yields larger probabilities, while
still ensuring that the probabilities do not exceed 1.

To obtain the normed conditional feature probability, we distinguish
whether the feature already appeared in conjunction with the formula. In
case it did, its probability is
\[P(f_j \mid l_i, \exists \vec{f'} \in F(l_i). f_j \in \vec{f'}) =
  \frac{\sum _ {\vec{f'} \in F(l_i)} 1 _ {\vec{f'}}(f_j)}{\left| F(l_i) \right|},\]
where \(1_A(x)\) denotes the indicator function that returns 1 if
\(x \in A\) and 0 otherwise. In case the feature \(f_j\) has never
appeared with the rule \(l_i\) before, we attribute it some minimal
probability with respect to all current features \(\vec f\) and all
currently applicable rules \(\vec l\):
\[P(f_j \mid l_i, \lnot \exists \vec{f'} \in F(l_i). f_j \in \vec{f'}) =
  \min _ {f_j \in \vec f,\,l_i \in \vec l,\, \exists \vec{f'} \in F(l_i). f_j \in \vec{f'}}
  P(f_j \mid l_i)\] The two definitions form a complete description of
the normed feature probability \(P(f_j \mid l_i)\).

\hypertarget{state-reward}{%
\subsection{State Reward}\label{state-reward}}

The state reward \(\rho(s)\) is evaluated for the final state \(s\) of a
random simulation. It estimates the likelihood of finding a proof from
any ancestor of the starting node of the random simulation. Therefore,
the state reward influences which regions of the Monte Carlo tree are
explored.

As the state reward is only calculated once per random simulation, it
can in practice be a function that is more expensive to calculate than,
say, the transition probability. A baseline state reward function
\(\rho_r\) returns random values between 0 and 1.

To estimate the \emph{discrimination} of a heuristic, i.e.~its ability
to distinguish nodes that lead to proofs from nodes that do not, we take
the ratio of the average rewards on the Monte Carlo tree branch leading
to a proof and the average rewards of all Monte Carlo tree nodes.

\hypertarget{beta-ratio}{%
\subsubsection{\texorpdfstring{\(\beta\)-ratio}{\textbackslash{}beta-ratio}}\label{beta-ratio}}

The \(\beta\)-ratio reward function considers the ratio of closed
\(\beta\)-children and all \(\beta\)-children in the tableau:
\[\rho _ \beta(s) = \frac{\left|\beta_c(s)\right|}{\left|\beta(s)\right|}.\]
This heuristic guarantees that for a closed tableau \(s\), the reward
\(\rho _ \beta(s)\) is 1.

\begin{example}
For state 4 in \autoref{fig:proof-search}, there are
five closed $\beta$-children and seven $\beta$-children in total.
Therefore, the reward $\rho _ \beta$ is $\frac{5}{7}.$
\end{example}

\hypertarget{formula-weight-reward}{%
\subsubsection{Formula Weight Reward}\label{formula-weight-reward}}

The formula weight reward heuristic calculates the average inverse
weight (i.e.~formula size) of all open \(\beta\)-children, encouraging
tableaux with smaller formulae. Furthermore, the heuristic gives higher
impact to formulae closer to the root, because the closer to the root a
formula is in the tableau, the more likely it is to be chosen in other
random simulations from the same starting node, therefore it is more
characteristic for the starting node. For that reason, the heuristic
weighs every inverse formula weight with the \emph{depth} of the formula
in the tableau, where the depth of a formula \(f\) in a tableau is
expressed as \(d(f)\). However, because rewards need to be normed
between 0 and 1, the depth needs to be normalised. For that purpose, we
introduce the concept of a \emph{normalisation function}.

\begin{definition}[Normalisation function]
A \emph{normalisation function} $N^u_l : [0, \infty) \to [u, l)$ with $l < u$
is strictly increasing and fulfils
$\lim_{x\to\infty} N^u_l(x) = u$ and $N^u_l(0) = l$.
\end{definition}

We choose the normalisation function
\(N^u_l(x) = u - \left(x + (u-l)^{-1}\right)^{-1}\). This allows us to
write the final formula weight function:
\[\rho_w(s) = \frac{1}{\left|\beta_o(s)\right|}
  \sum _ {c \in \beta_o(s)} \frac{1}{\left|c\right|} N_l^1 (d(c)),
\] where \(l>0\) is a constant that determines the impact of formula
depth. For example, when \(l = 1\), then depth has no influence
whatsoever, whereas \(l \approx 0\) gives hardly any weight to formulae
close to the root. In this particular \(\rho_w\), we use the arithmetic
mean, but we have also experimented with geometric and harmonic means as
well as the minimum.

\begin{example}
The open $\beta$-children $r$ and $t$ in state 4 of \autoref{fig:proof-search}
are at depth 1 and 2, respectively.
Therefore, the formula weight reward of the tableau is the mean of
$\frac{1}{\left|r\right|}N_l^1(1)$ and
$\frac{1}{\left|t\right|}N_l^1(2)$.
\end{example}

This heuristic is based on similar ideas as the \emph{pick-given ratio}
popularised by Otter \cite{schulz2002-brainiac}.

\hypertarget{refutability}{%
\subsubsection{Machine-Learnt Refutability
Estimate}\label{refutability}}

The \emph{refutability} of a tableau \(s\) can be estimated with
knowledge how often open \(\beta\)-children of \(s\) were successfully
refuted in previous proofs.

We call a formula refuted when all branches on which it lies are closed.
A formula is unsuccessfully refuted if it is present in the tableau, but
lies on at least one open branch. Note that refuted \(\beta\)-children
are always closed (as defined in \autoref{tableau}), but closed
\(\beta\)-children are not necessarily refuted.

\begin{example}
The $\beta$-child $q$ in state 4 of \autoref{fig:proof-search} is closed,
but not refuted.
\end{example}

When statistics about previous refutations of formulae are available, we
use them to estimate the refutability of formulae in the current proof
search, similarly to \cite{faerber2016-satallax}. Let \(p(f)\) be the
number of successful and \(n(f)\) the number of unsuccessful refutations
of a formula \(f\). Then the irrefutability ratio of \(f\) is
\(\frac{n(f)}{p(f)+n(f)}\).

We want the irrefutability ratio to have an effect proportional to the
amount of information available about previous refutation attempts.
Consider the case for a formula \(f\) where \(p(f) = 0\) and
\(n(f) = 1\). The irrefutability ratio of \(f\) then is 100\%, but
because we have information about only a single refutation attempt, we
want to attribute less meaning to it compared to, say, a formula where
\(p(f) = 0\) and \(n(f) = 1000\). To achieve this, we weigh the
irrefutability with \(N_u^l (v(p(f)+n(f)))\), where \(v \geq 0\),
\(u \geq 0\) and \(l \leq 1\) are constants. This term reflects the
\emph{confidence} in the irrefutability ratio. \(v\) determines how fast
we gain confidence, \(u\) is the minimal and \(l\) is the maximal
confidence.

The estimated refutability of the formula \(f\) then is the opposite of
its confidence-weighted irrefutability:
\[1 - N_u^l (v(p(f)+n(f))) \frac{n(f)}{p(f)+n(f)}\]

The machine-learnt refutability estimate of a whole tableau is the mean
of estimated refutabilities of the tableau's open \(\beta\)-children.

\begin{example}
The open $\beta$-children in state 4 of \autoref{fig:proof-search}
are $t$ and $r$.
Assume that $p(t) = 222$, $n(t) = 115$, $p(r) = 62$, and $n(r) = 553$.
Then the machine-learnt refutability estimate of the tableau
is the mean of
$1 - N_u^l (v \cdot 337) \frac{115}{337}$ and
$1 - N_u^l (v \cdot 615) \frac{553}{615}$.
In case we have total confidence in the statistics (e.g. by setting $u = l = 1$)
and use the arithmetic mean, the resulting refutability estimate is 0.38.
\end{example}

\hypertarget{beta-min-expansion}{%
\subsection{\texorpdfstring{\(\beta\)-minimal Expansion
Policies}{\textbackslash{}beta-minimal Expansion Policies}}\label{beta-min-expansion}}

The default expansion policy in \autoref{mcsf} creates new nodes in the
Monte Carlo tree from the first state \(s_1\) of a random simulation
\(\left[s_1, \dots, s_D \right]\). This can be counterproductive in
cases where the random simulation closes a subtree, but fails to find a
proof in the end. In that case, keeping the successful part of the proof
attempt, i.e.~the closed subtree, can accelerate proof search.

This motivates \(\beta\)-minimal expansion policies, where new nodes are
created not from the first state of a simulation, but from some state
minimising a function related to \(\beta\)-children.

The first policy is the \(\beta\)-child expansion policy, which chooses
the state with fewest open \(\beta\)-children, i.e.,
\(\min _ i \left| \beta_o(s_i) \right|\).

The second policy is the \(\beta\)-parent expansion policy, which
chooses the state with fewest parents of open \(\beta\)-children, i.e.
\(\min _ i \left| \bigcup _ {o \in \beta_o(s_i)} p(o) \right|\), where
\(p(s)\) denotes the parent of a node \(s\).

Similarly to restricted backtracking \cite{otten2010-cut}, the
\(\beta\)-minimal expansion policies lose completeness, but can in
practice perform significantly better than complete strategies.

\begin{example}
In the proof search in \autoref{fig:proof-search}, the proof attempt failed.
We assume that the proof search started from
a Monte Carlo node $n$ containing state 1.
The default expansion policy would add a new node corresponding to state 2
as child tree node of $n$ to the Monte Carlo tree.
However, this would discard the closed subtree found in state 3.
In contrast, the $\beta$-child expansion policy compares
the open $\beta$-children in all successor states of state 1:
State 2 has three open $\beta$-children ($s$, $q$, and $r$),
state 3 has two ($q$ and $r$) and state 4 has two as well ($t$ and $r$).
State 3 and 4 are therefore minimal, in which case
the first of them (i.e. state 3) is used as state for
a new Monte Carlo leaf node that is added as child tree node of $n$
to the Monte Carlo tree.
\end{example}

\hypertarget{implementation}{%
\section{Implementation}\label{implementation}}

We implemented the proposed Monte Carlo Tableau calculus in the OCaml
version \cite{kaliszyk2015-holcop} of leanCoP
\cite{otten2008-leancop}. The implementation and experimental data are
available at:
\url{http://cl-informatik.uibk.ac.at/users/mfaerber/cade-26.html}. In
the rest of this paper, we refer to the OCaml version of leanCoP as
leanCoP.

Monte Carlo proof search can be used to \emph{advise} a \emph{base
prover}: The proof search is conducted by a base prover such as leanCoP.
When the base prover has a choice between different applicable proof
rules, it starts the advisor, i.e.~Monte Carlo proof search, which
returns after a certain number of iterations an order on the proof rules
to be tried by the base prover. This order is based on the average Monte
Carlo rewards achieved for each rule. Furthermore, when Monte Carlo
proof search finds proofs while establishing the proof rule order, the
proofs are used directly by the base prover. In the extreme case, when
setting the number of Monte Carlo iterations to \(\infty\), the whole
proof search is done by Monte Carlo proof search and the base prover is
only responsible for starting it and printing the proof. We refer to our
implementation of Monte Carlo proof search as advisor for leanCoP as
\emph{Monte Carlo Prover}.

In contrast to leanCoP, Monte Carlo proof search does not require
iterative deepening. Instead, an important parameter is the simulation
depth \(D\) as shown in \autoref{mcsf}, which determines the length of
random simulations.

\begin{figure}
  \subfloat[Iterative deepening without restricted backtracking.]{%
    \includegraphics[width=100pt]{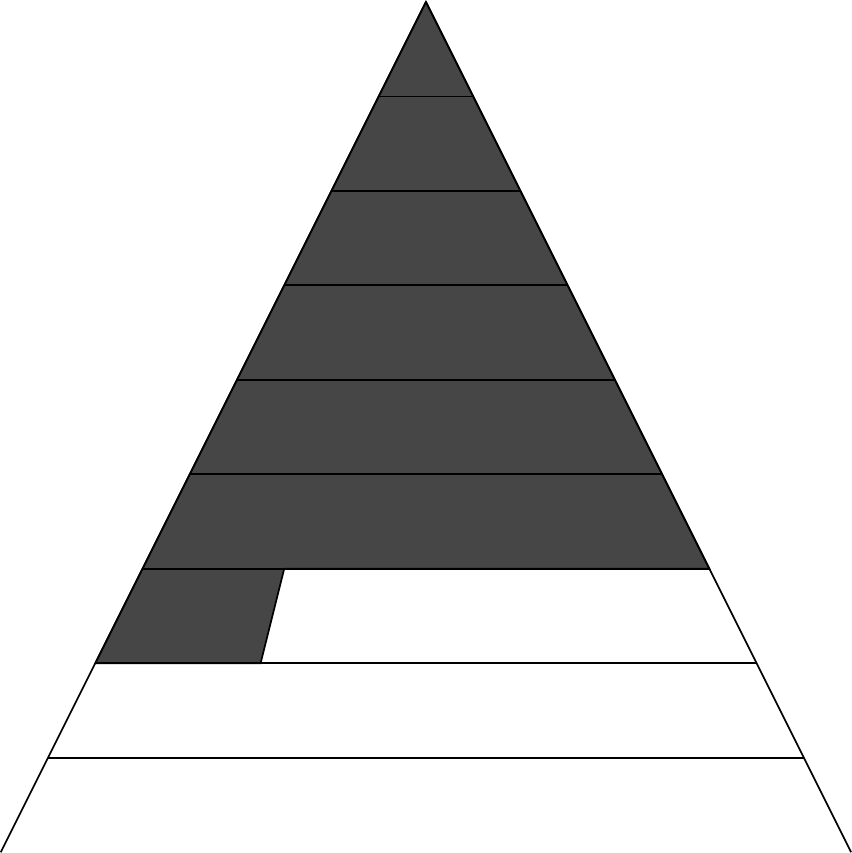}
  }
  \hfill
  \subfloat[Iterative deepening with restricted backtracking.]{%
    \includegraphics[width=100pt]{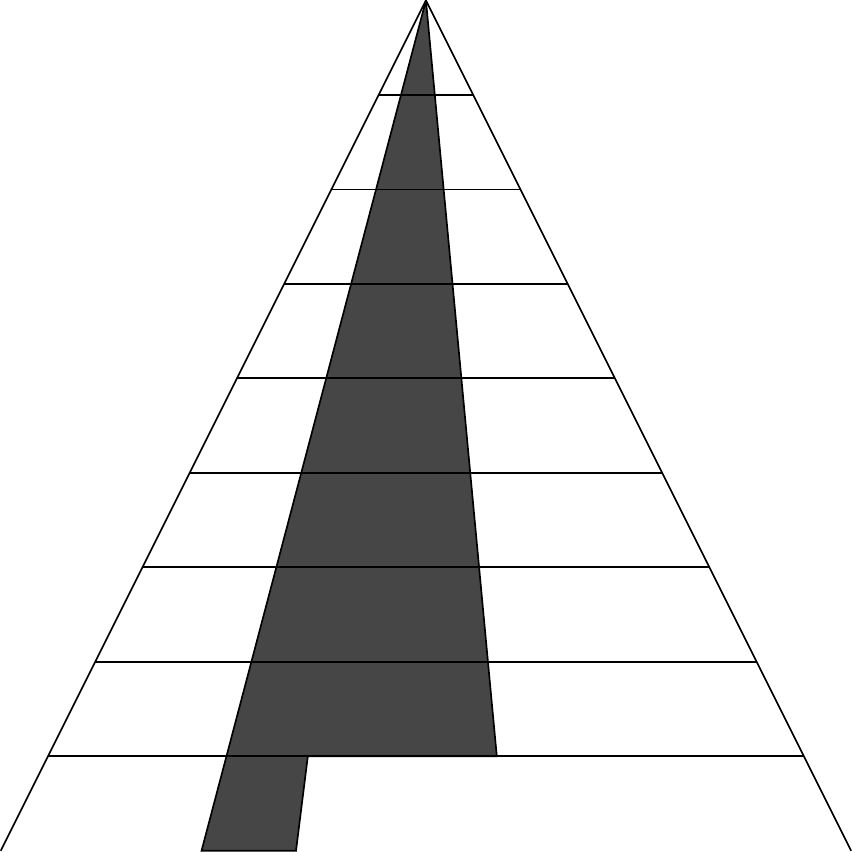}
  }
  \hfill
  \subfloat[Monte Carlo.]{%
    \includegraphics[width=100pt]{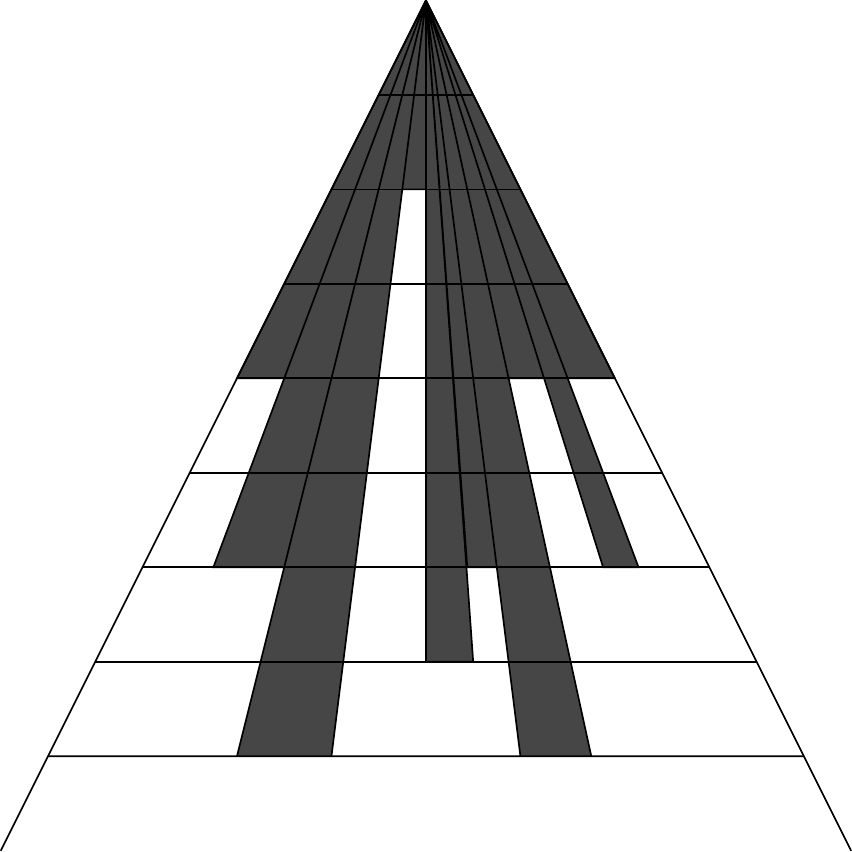}
  }
  \caption{The two main leanCoP strategies compared with Monte Carlo proof search.}
  \label{fig:leancop-montecop}
\end{figure}

leanCoP is equipped with a set of strategies, where each strategy
consists of a set of options, such as whether to use definitional
clausal normal form. A strategy schedule tries different strategies for
a defined amount of time until a strategy succeeds. One of the most
influential developments in leanCoP was restricted backtracking
\cite{otten2010-cut}, which discards other possibilities to close a
subtree once it has been closed. See \autoref{fig:leancop-montecop} for
a comparison of the complete strategy with the restricted backtracking
strategy, as well as an illustration of a Monte Carlo search.

In the next section, we evaluate how well our Monte Carlo prover
performs in comparison to single leanCoP strategies.

\hypertarget{evaluation}{%
\section{Evaluation}\label{evaluation}}

In this section, we evaluate the Monte Carlo prover described in
\autoref{implementation}. We first describe the dataset and the
evaluation parameters. Then we evaluate the different heuristics given
in \autoref{tableau-heuristics}, as well as the influence of several
numeric parameters. Finally, we show our best obtained Monte Carlo
configuration and compare it to leanCoP.

\hypertarget{experimental-setup}{%
\subsubsection*{Experimental Setup}\label{experimental-setup}}
\addcontentsline{toc}{subsubsection}{Experimental Setup}

We used the bushy version of the MPTP2078 dataset
\cite{alama2014-premsel}, which is particularly valuable for our
machine learning algorithms as it provides consistent symbols over all
problems. To generate training data for the machine learning heuristics,
we ran leanCoP for 60s on all the MPTP2078 problems, using a strategy
schedule with three strategies, including a restricted backtracking and
a complete strategy. The outcome of the training runs were formula
usability data for the Naive Bayes heuristic in \autoref{naive-bayes} as
well as formula refutability data for the heuristic in
\autoref{refutability}.

For the main evaluation, we used definitional clausification and a
timeout of 10s per problem for both leanCoP and the Monte Carlo prover,
where the 10s timeout is also used for the MPTP2078 evaluation in
\cite{kaliszyk2015-holcop}. In that setting, leanCoP solves 509
problems with restricted backtracking and 388 without, the union being
562 problems. In the remainder of this paper, leanCoP refers to the
restricted backtracking strategy of leanCoP.

For the Monte Carlo prover, we used the following initial parameters:

\begin{itemize}
\tightlist
\item
  Maximal simulation depth \(D\): 50
\item
  Exploration constant \(C_p\): 1 (see \autoref{mcsf})
\item
  Transition probability: \(\beta\)-size (see \autoref{beta-size})
\item
  State reward: \(\beta\)-ratio (see \autoref{beta-ratio})
\item
  Depth attenuation for formula weight reward: 0 (see
  \autoref{formula-weight-reward})
\item
  Refutability mean: \(\min\) (see \autoref{refutability})
\item
  Refutability confidence velocity: 1 (see \autoref{refutability})
\item
  Minimal/maximal refutability confidence: 0/1 (see
  \autoref{refutability})
\item
  Expansion policy: \(\beta\)-child expansion policy (see
  \autoref{beta-min-expansion})
\end{itemize}

\hypertarget{heuristics-influence}{%
\subsubsection*{Heuristics Influence}\label{heuristics-influence}}
\addcontentsline{toc}{subsubsection}{Heuristics Influence}

We evaluated the Monte Carlo prover with a set of configurations where
each configuration deviates by one heuristic from the initial
parameters. For every configuration, we collected the set of solved
problems. Furthermore, we collected the problems solved by all Monte
Carlo configurations, amounting to 196 problems. On these problems, for
all Monte Carlo configurations, we evaluated the average number of MCTS
iterations and MCTS simulation steps, as well as the average reward
discrimination; see \autoref{montecarlo-results}.

The machine-learnt reward heuristic performs best, with a very good
discrimination rate of 2.30. Surprisingly, the random reward heuristic
solves only three problems less, despite its worse discrimination.

The Bayesian transition probability shows very poor performance. The
\(\beta\)-size heuristic is the winner for transition probability.

The \(\beta\)-parent expansion policy outperforms the default expansion
policy by 20 problems, i.e.~6\%.

\begin{table}
\caption{Comparison of Monte Carlo heuristics.
Iterations, simulation steps and discrimination ratio are averages
on the 196 problems solved by all configurations.}
\label{montecarlo-results}

\begin{tabular}[c]{lrrrr}
\toprule
Configuration & Iterations & Sim. steps & Discr. & Solved\tabularnewline
\midrule
Base & 116.46 & 1389.82 & 1.37 & 332\tabularnewline
Random reward & 104.88 & 1167.98 & 1.19 & 364\tabularnewline
Formula weight reward & 108.13 & 1268.88 & 1.12 & 334\tabularnewline
ML reward & 108.52 & 1151.61 & \textbf{2.30} & \textbf{367}\tabularnewline
Bayes \(P\) & 528.39 & 8014.03 & 1.35 & 248\tabularnewline
Constant \(P\) & 949.62 & 17539.59 & 1.31 & 237\tabularnewline
\(\beta\)-parent exp. & 224.72 & 2769.12 & 1.40 & 348\tabularnewline
Default exp. & 371.81 & 4793.58 & 1.38 & 328\tabularnewline
\bottomrule
\end{tabular}
\end{table}

\hypertarget{parameter-influence}{%
\subsubsection*{Parameter Influence}\label{parameter-influence}}
\addcontentsline{toc}{subsubsection}{Parameter Influence}

We identified three numeric parameters to be highly influential for
proof search; namely the simulation depth \(D\), the exploration
constant \(C_p\), and the maximal number of MCTS iterations per base
prover step. We evaluated a large range of values for these parameters,
keeping the remaining parameters fixed to the standard values. The
results are shown in \autoref{fig:param-discussion}.

We achieve the highest performance of the Monte Carlo prover when using
it as an advisor for a base prover. From \autoref{fig:maxiters}, it
becomes clear that the Monte Carlo prover is most useful when given
between 20 and 40 iterations per base prover step. Below that mark, the
reward estimates are too imprecise, and above that mark, the reward
precision increases only marginally, compared to the time spent in the
MCTS prover.

The higher the maximal simulation depth \(D\) (see
\autoref{fig:simdepth}), the more time the prover spends looking for
proofs at less promising higher depths. \autoref{fig:simdepth-ss} shows
that the average number of simulation steps decreases with increasing
\(D\). This indicates that at higher simulation depths, the
computational effort to calculate the set of possible steps increases.

\autoref{fig:exploration} shows the number of solved problems for the
\(\beta\)-ratio and the machine-learnt state evaluation heuristics as
function of the exploration constant \(C_p\). For a good state reward
heuristic, one expects in such a graph a local optimum, where
exploration and exploitation combine each other best. As one can see,
this is given for the machine-learnt heuristic at \(C_p \approx 0.75\),
whereas the curve for the \(\beta\)-ratio heuristic does not expose such
an optimum.

\begin{figure}[ht!]
  \subfloat[Maximal number of MCTS iterations.]{\label{fig:maxiters}%
    \begin{tikzpicture}[scale=0.65]
    \begin{axis}
    [ legend pos=south east
    , xlabel=MCTS iterations
    , ylabel=Problems solved
    ]
    \addplot [mark=none] table {data/montecop-170209-maxiters};
    \end{axis}
    \end{tikzpicture}
  }
  \hfill
  \subfloat[Simulation depth $D$.]{\label{fig:simdepth}%
    \begin{tikzpicture}[scale=0.65]
    \begin{axis}
    [ legend pos=south east
    , xlabel=Simulation depth
    , ylabel=Problems solved
    ]
    \addplot [mark=none] table {data/montecop-170209-simdepth};
    \end{axis}
    \end{tikzpicture}
  }
  \hfill
  \subfloat[Simulation steps / Simulation depth.]{\label{fig:simdepth-ss}%
    \begin{tikzpicture}[scale=0.65]
    \begin{axis}
    [ legend pos=south east
    , xlabel=Simulation depth
    , ylabel=Simulation steps
    ]
    \addplot [mark=none] table {data/montecop-170209-simdepth-simsteps};
    \end{axis}
    \end{tikzpicture}
  }
  \hfill
  \subfloat[Exploration constant $C_p$.]{\label{fig:exploration}%
    \begin{tikzpicture}[scale=0.65]
    \begin{axis}
    [ legend pos=south east
    , xlabel=$C_p$
    , ylabel=Problems solved
    ]
    \addplot+ [mark=none] table {data/montecop-170209-mlrew-exploration};
    \addplot+ [dashed,mark=none] table {data/montecop-170209-exploration};
    \legend{Machine-learnt heuristic,$\beta$-ratio}
    \end{axis}
    \end{tikzpicture}
  }

\caption{Parameter influence.}
\label{fig:param-discussion}
\end{figure}

\hypertarget{best-found-monte-carlo-configuration}{%
\subsubsection*{Best Found Monte Carlo
Configuration}\label{best-found-monte-carlo-configuration}}
\addcontentsline{toc}{subsubsection}{Best Found Monte Carlo
Configuration}

Our best found configuration MC\(^+\) for the Monte Carlo prover uses
the arithmetic mean for the ML reward, a maximal number of 27 MCTS
iterations and a simulation depth of 20. Interestingly, is has a
discrimination ratio of only 1.07, which suggests that a high
discrimination ratio indicates good performance, but is not absolutely
necessary to achieve it.

MC\(^+\) performs on average 902 times more inferences in MCTS than in
the base prover. Furthermore, for the problems solved both by leanCoP
and by MC\(^+\), leanCoP takes on average 21698 inferences, while
MC\(^+\) takes 20243 inferences (sum of base prover + MCTS inferences).

MC\(^+\) solves 538 problems, compared to 509 by leanCoP. Of the 538
problems, 90 problems were previously not solved by leanCoP. The union
of MC\(^+\) and leanCoP solves 599 problems, compared to 531 problems
solved by leanCoP with a timeout of 20s. That means that we solve 12.8\%
more problems. Furthermore, MC\(^+\) proves more problems than leanCoP
when given only half the time.

\begin{longtable}[]{@{}lrr@{}}
\toprule
Prover & Timeout {[}s{]} & Solved problems\tabularnewline
\midrule
\endhead
leanCoP & 10 & 509\tabularnewline
MC\(^+\) & 10 & 538\tabularnewline
leanCoP + MC\(^+\) & 10+10 & 599\tabularnewline
leanCoP & 20 & 531\tabularnewline
\bottomrule
\end{longtable}

\hypertarget{conclusion}{%
\section{Conclusion}\label{conclusion}}

We have proposed a combination of Monte Carlo Tree Search and tableau
automated theorem proving. MCTS provides a theoretically founded
fine-grained mechanism to control the search space of tableau-based
theorem provers based on random sampling and state evaluation
heuristics, which might eventually even replace iterative deepening. We
have shown that a fast rollout policy combined with a machine-learnt
state evaluation heuristic and a custom expansion policy produce the
best results. The strength of the current system has turned out to be
its function as advisor for existing provers, demonstrated by our
integration into leanCoP. This opens a wide space of future work,
profiting from the ongoing research in MCTS; examples include
self-updating reward heuristics, adaptive simulation depths, automatic
parameter tuning, and different characterisations of tableau search or
expansion policies such as AMAF to produce more shallow Monte Carlo
trees. Furthermore, identifying controversial choices in the base prover
would allow using the Monte Carlo prover as advisor more efficiently.
Finally, neural networks could be used as state reward heuristics.

\hypertarget{acknowledgements}{%
\subsubsection*{Acknowledgements}\label{acknowledgements}}
\addcontentsline{toc}{subsubsection}{Acknowledgements}

We thank the anonymous CPP and CADE referees for their valuable comments
on previous versions of this paper. This work has been supported by the
Austrian Science Fund (FWF) grant P26201 and the European Research
Council (ERC) grants no. 649043 \emph{AI4REASON} and no. 714034
\emph{SMART}.


\end{document}